\documentclass[twoside]{article}
\usepackage{fleqn}
\usepackage{espcrc2}
\usepackage{graphicx}
\usepackage{epsfig}
\usepackage{amsmath}
\usepackage[figuresright]{rotating}

\begin{document}

%\vspace*{-2cm}
%\begin{flushleft}
%DESY 04-095 \\
%SFB/CPP-04-21
%\end{flushleft}
% declarations for front matter
\title{%
A complete set of scalar master integrals for massive 2-loop Bhabha
scattering: where we are 
\thanks{Work supported in part 
  by European's 5-th Framework under contract HPRN--CT--2000--00149 Physics at
  Colliders, 
  by TMR, EC-Contract No. HPRN-CT-2002-00311 (EURIDICE), 
  by Deutsche Forschungsgemeinschaft under contract SFB/TR 9--03, 
  and by the Polish State Committee for Scientific Research (KBN)
  under contract Nos. 2P03B01025 and 1P03B00927.%\\
} }

\author{M. Czakon\address{Deutsches Elektronen-Synchrotron, DESY
    Zeuthen, %\\
   Platanenallee 6, 15738 Zeuthen, Germany}%
\address{Institute of Physics, University of Silesia, %\\
   ul. Uniwersytecka 4, 40007 Katowice, Poland},
        J. Gluza$^{\text{ab}}$
        and
        T. Riemann$^{\text{a}}$
}

\begin{abstract}
We define a complete set of scalar master integrals (MIs) for 
massive 2-loop QED Bhabha scattering. Among others, there are 
thirty three 2-loop box type MIs.
Five of them  have been published in (semi-)analytical form, 
one is determined here, the rest remains to be calculated.
Further, the last four so far unknown 2-loop 3-point MIs are identified
and also computed here.
\end{abstract}

\maketitle

%-----------------------------------------------------------------------
\section{\label{sec-intro}INTRODUCTION}
%-----------------------------------------------------------------------
%
The determination of luminosity at high energy colliders is 
a backbone linking theoretical calculations with experimental data. 
Bhabha scattering,
%\begin{eqnarray}
$e^+e^- \to e^+e^-(n \gamma),$ is the process that allows to accomplish 
this task. 
%\end{eqnarray}
%is an important process because it
So far the complete NNLO corrections in
massive QED have not been computed. There are several reasons 
to fill this gap. First, the future experimental accuracy is expected 
to be of the order of $10^{-4}$ 
\cite{Lohmann:2004nn}. 
%\cite{Lohmann:2004nn,Battaglia:2001dg}. 
Obviously, 
the  theoretical error should be smaller.
Second, the current  Monte Carlo 
generators for real photon radiation assume 
massive external fermions. 
Analytical results of NNLO corrections 
in massive QED  are clearly needed there for consistency.
There are also 
technical reasons to perform the calculation.  
First, as argued 
in \cite{Bern:2000ie}, 2-loop QED results are a useful testing 
ground for 2-loop QCD physics containing more than one kinematic 
invariant. Second, the full analytic results can be easily 
approximated to various kinematic regions;
this is not possible for massless results.
Finally, 
%One may search for New Physics due to four fermion operators of the
%type $e^+e^-e^+e^-$, and in the Standard Model it is an ideal base for
%luminosity-monitoring.
%In the very forward region, the reaction is dominated by pure photonic
%contributions due to the peaking of the massless $t$-channel
%propagator.
%and theoretical predictions for low angle Bhabha scattering should include
%complete 1-loop corrections in the Standard Model plus 2-loop
%corrections in QED.
%From a purely technical point of view, 
the complete calculation of
virtual 2-loop corrections in massive QED is a prelude to the 
determination of analogous effects within the Standard Model.

\vspace{-.75cm}
\begin{figure}[htb]
\epsfig{file=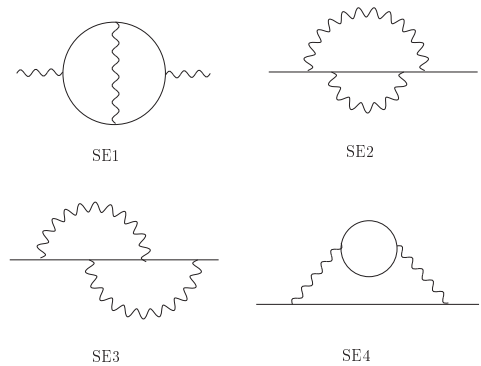, width=6.5cm}
\vspace{-1cm}
\caption{The 2-loop self energy diagrams.}
\label{SEprot}
\end{figure}
%\vspace{-0.5cm}
In this contribution, we report on first results of our calculation of
virtual 2-loop corrections in massive QED.
We have determined a complete set of scalar master integrals to be
calculated for the problem at hand, including 33 2-loop box master
integrals\footnote{%
%The complete set of  results, 
More detailed information, including 
the MIs computed so far, can be found at the webpage {\bf bhabha.html} in
\cite{web-masters:2004nn}.}. 
%\newline
%\begin{verbatim}
%{\bf http://www-zeuthen.desy.de/\linebreak[2]theory/\linebreak[2]research/\linebreak[2]Bhabha.html}.}.
%\end{verbatim}}. 
%A comprehensive, but not complete study of 2-loop 3-point master
%integrals was given in 
Subsequently, we have calculated all 2- and 3-point MIs, 
four of which were still lacking in the literature.
We also started the calculation of boxes.

%An overview will be given in Section \ref{sec-topo}.
%Sections \ref{sec-ver} and \ref{sec-box} include both  new results
%for the last 3-point MIs which we need for Bhabha scattering and 
%calculation of a new box MI.
%Section  \ref{sec-cross} contains some cross checks, 
%Section \ref{sec-sum} includes the Conclusions/Summary.

%-----------------------------------------------------------------------
\section{\label{sec-topo}2--LOOP DIAGRAMS AND IDEN\-TI\-FI\-CATION OF
  MIs} 
\begin{figure}[htb]
\epsfig{file=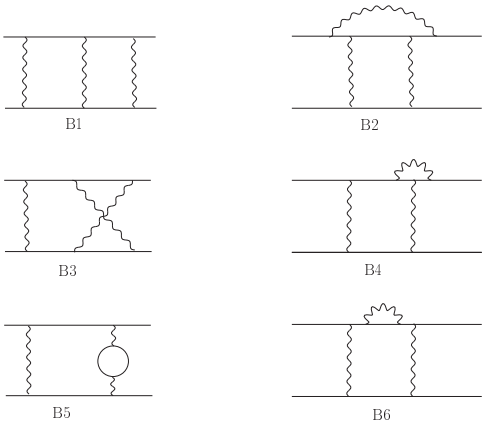, width=6.5cm}
\vspace*{-1cm}
\caption{The 2-loop box diagrams.}
\label{LL2}
\end{figure}

%  MASTER INTEGRALS} 
%-----------------------------------------------------------------------
%
In Figs. \ref{SEprot}
%,  \ref{beta} and Fig. 
to \ref{LL2} we show the set of 2-loop self energy, vertex
and box diagrams for which the scalar MIs are
needed.
%\footnote{Diagrams SE2-SE4 
%are needed for the renormalization of external electron lines.}.
%It is easy to note that some of drawn prototypes are actually
%also  subprototypes, e.g. SE2 and SE4 can be obtained by deleting 
%appropriate lines in SE3, the same is true for V4,V5 and B4,B5.
%It means that we have 4 basic prototypes for boxes. However, we left 
%them for ... reasons, let us note that B5 have been already fully 
%worked out in \cite{Bonciani:2003te}}.  

\vspace{-.5cm}
\begin{figure}[htb]
\epsfig{file=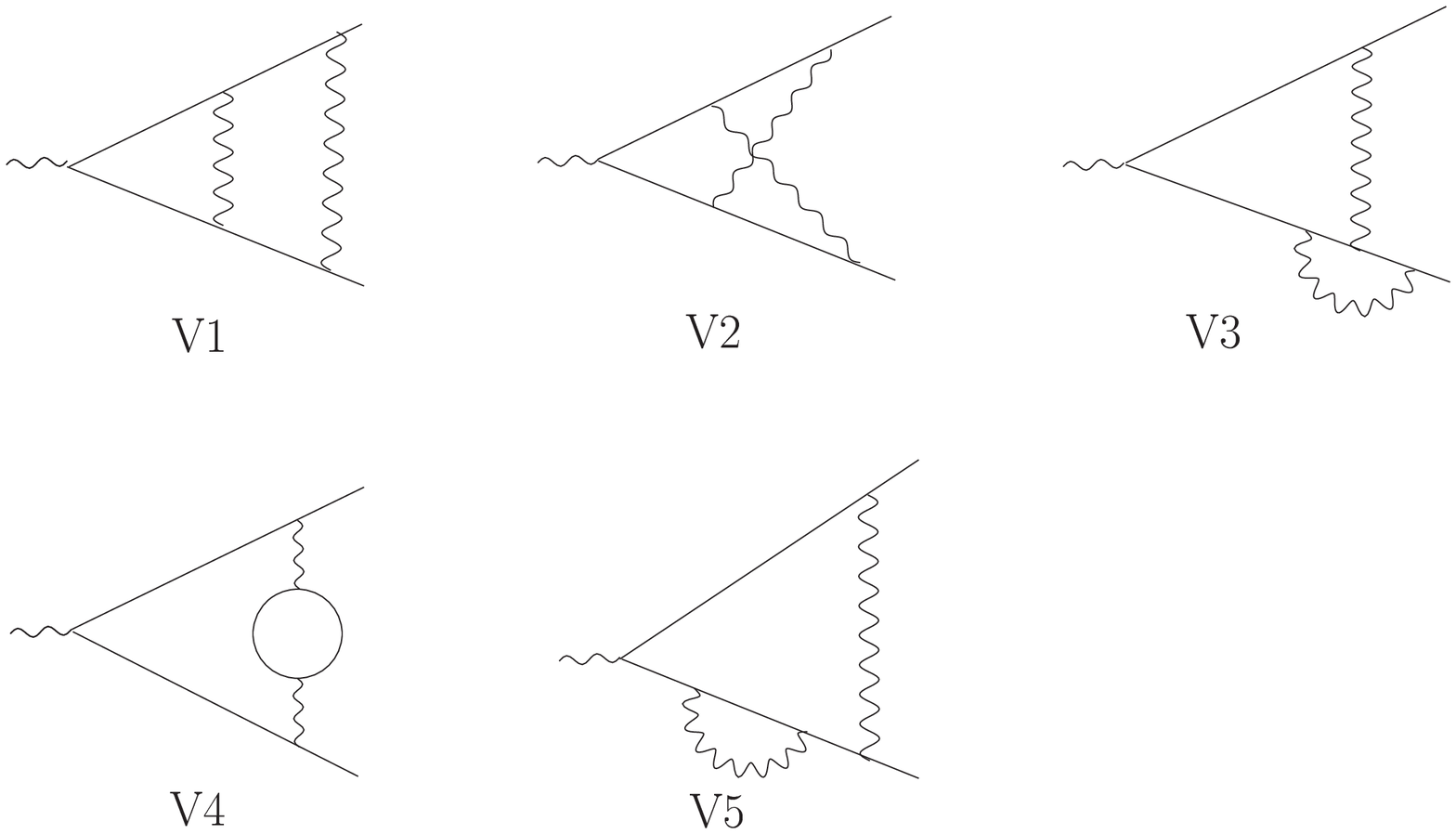, width=7.5cm}
\vspace*{-1.0cm}
\caption{The 2-loop vertex diagrams.}
\label{beta}
\end{figure}
\vspace{-1.5cm}
\vspace*{-.5cm}

\begin{figure}[bht]
\epsfig{file=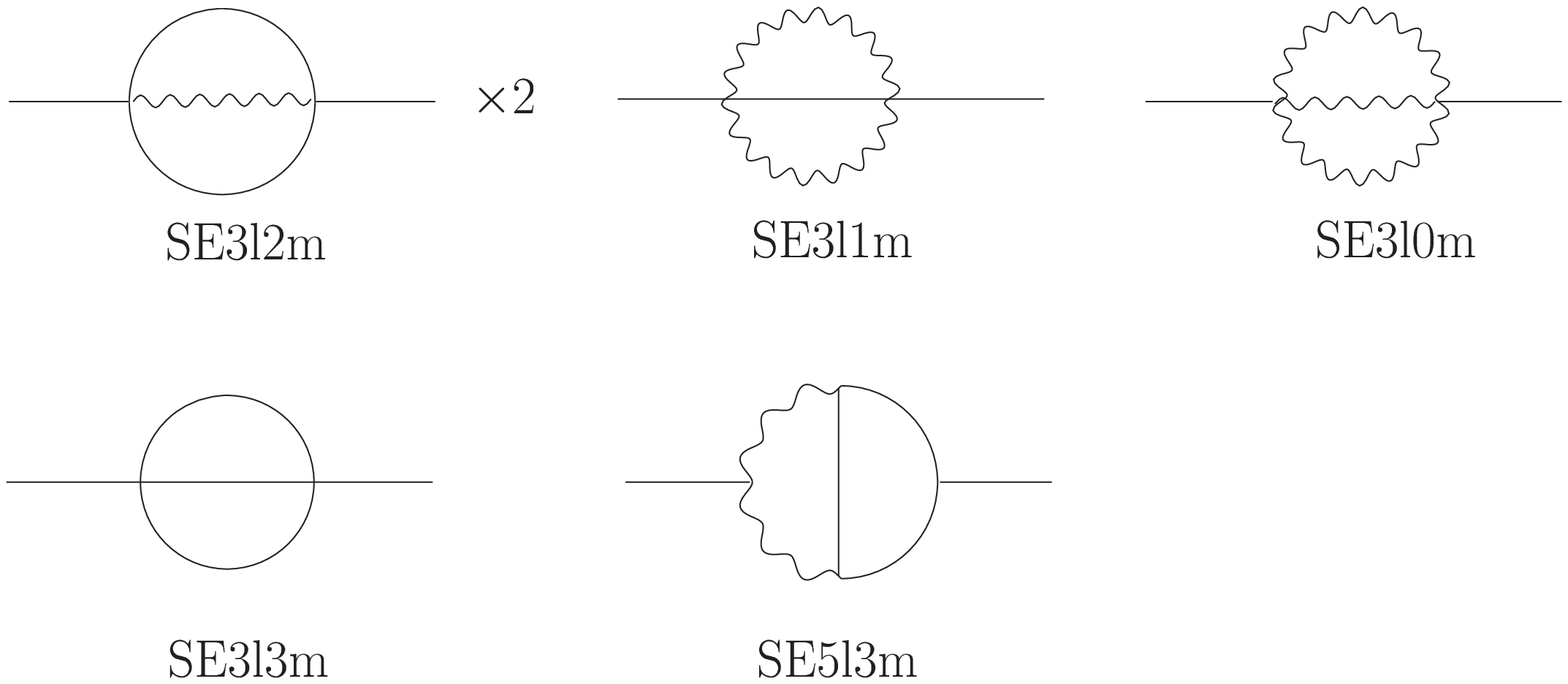, width=7.5cm}
\vspace{-1.1cm}
\caption{The six 2-loop 2-point MIs. The multiplication factor stands for 
the number of masters for the given diagram when different from 1.}
\label{LL7}
\end{figure}

\begin{figure}[ht]
\epsfig{file=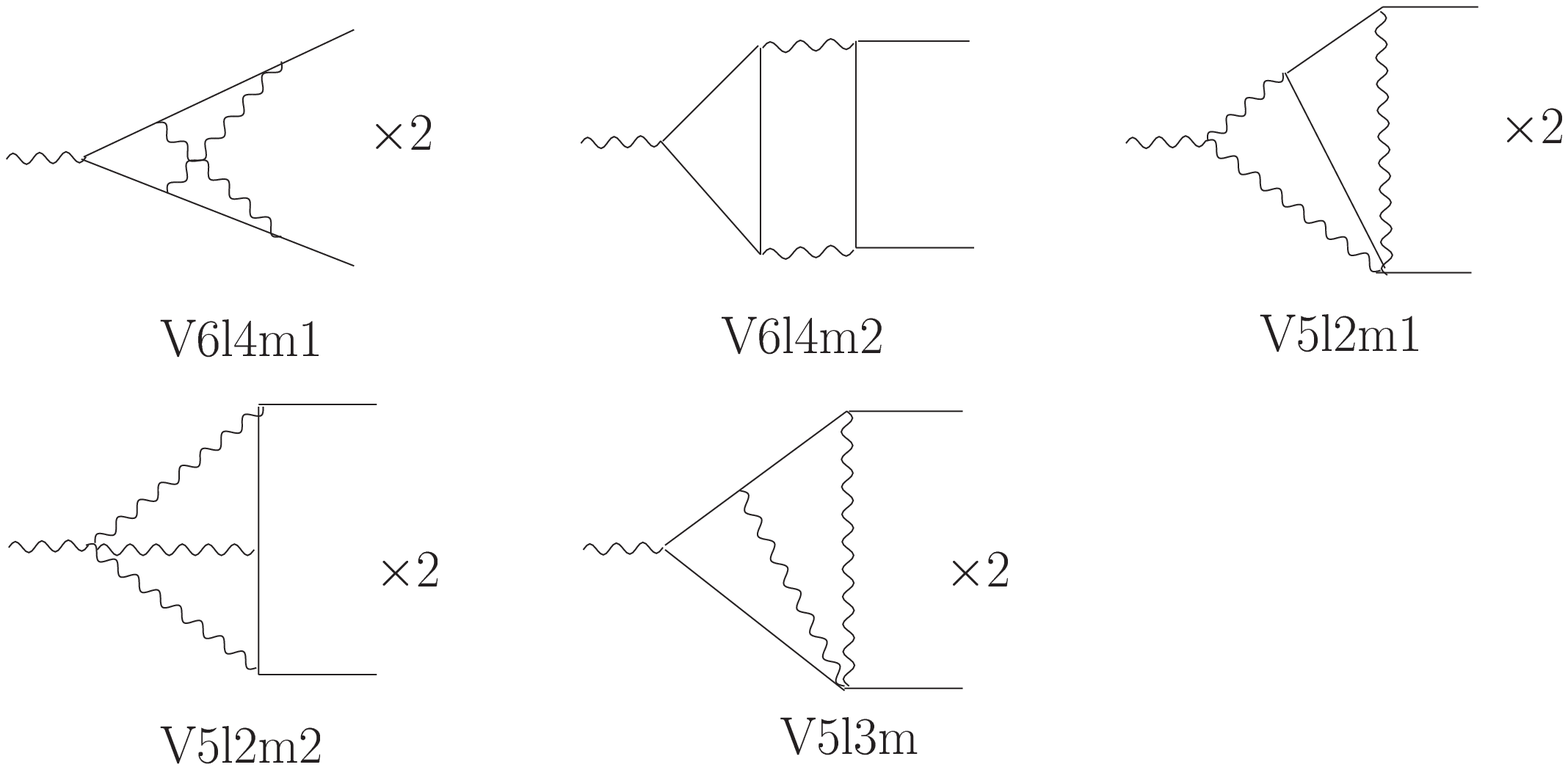, width=7.5cm}
%\vspace{1.5cm}
\epsfig{file=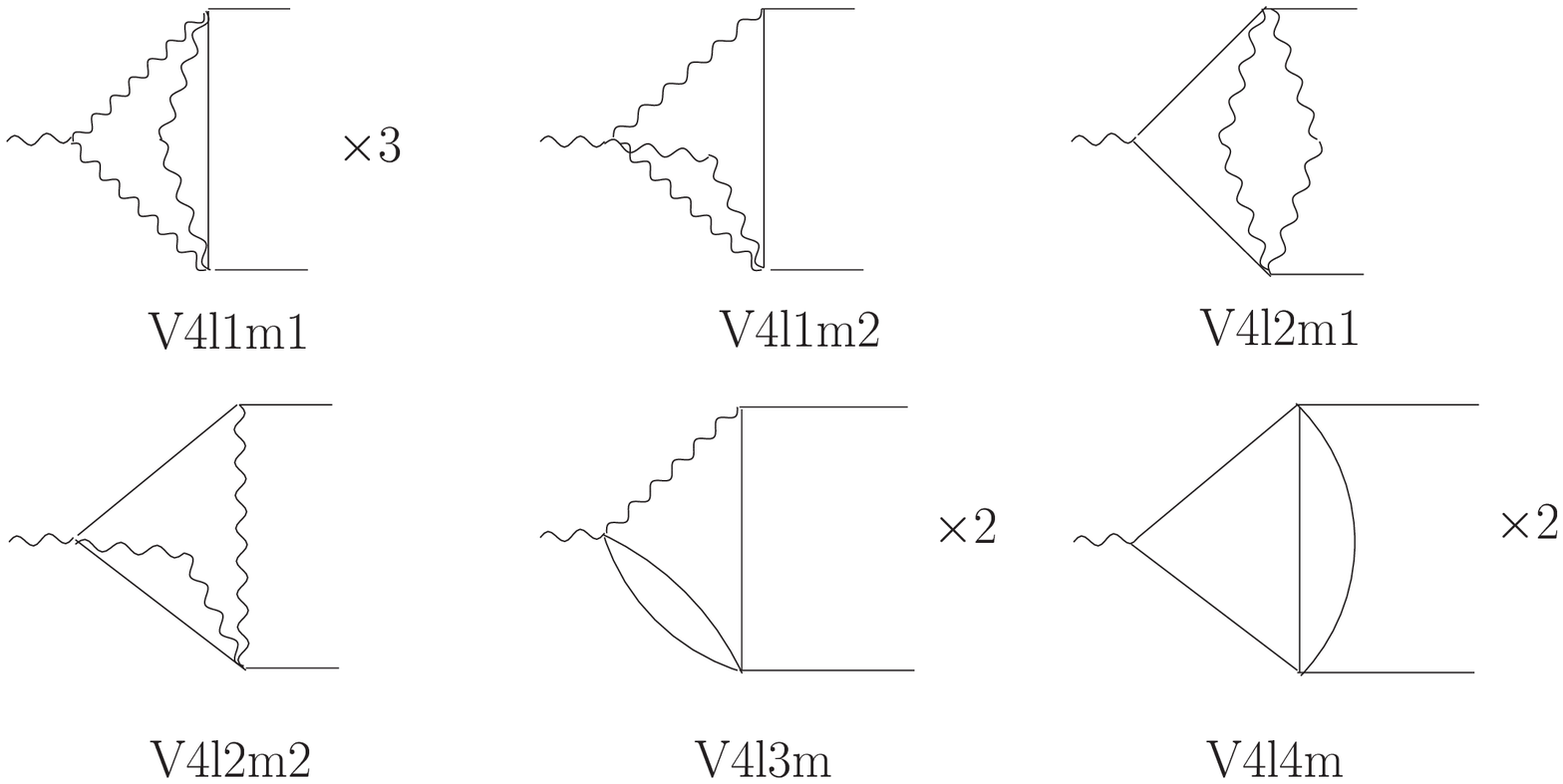, width=7.5cm}
\vspace{-1.cm}
\caption{The nineteen 2-loop 3-point MIs (the multiplication factors 
as in Fig. \ref{LL7}.)}
\label{beta2}
\end{figure}

\begin{figure}[ht]
\epsfig{file=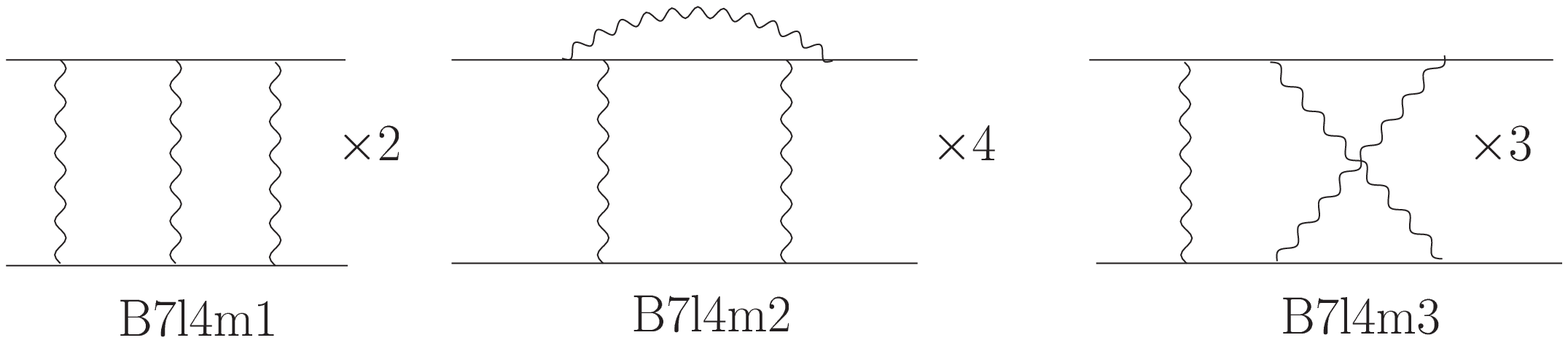, width=7.5cm}
\epsfig{file=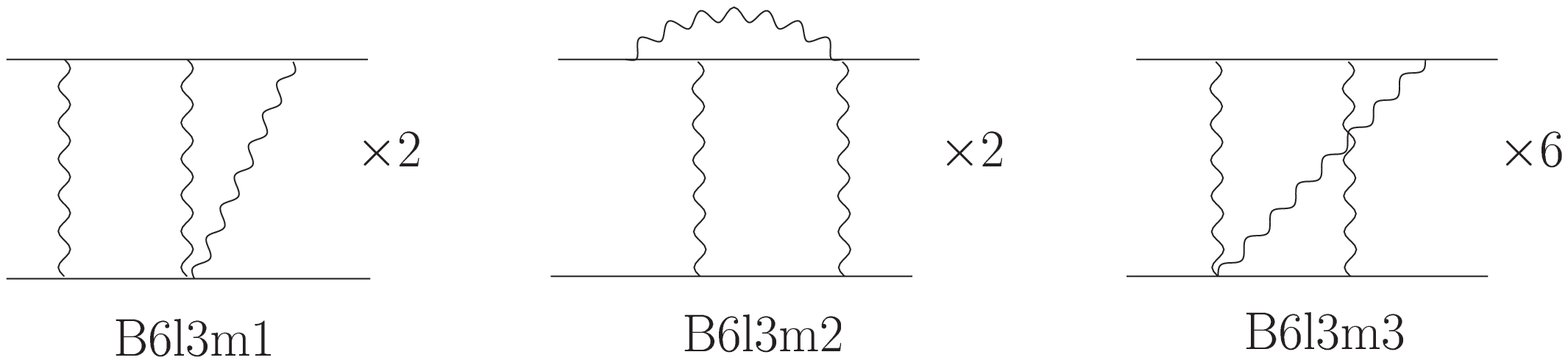, width=7.5cm}
\epsfig{file=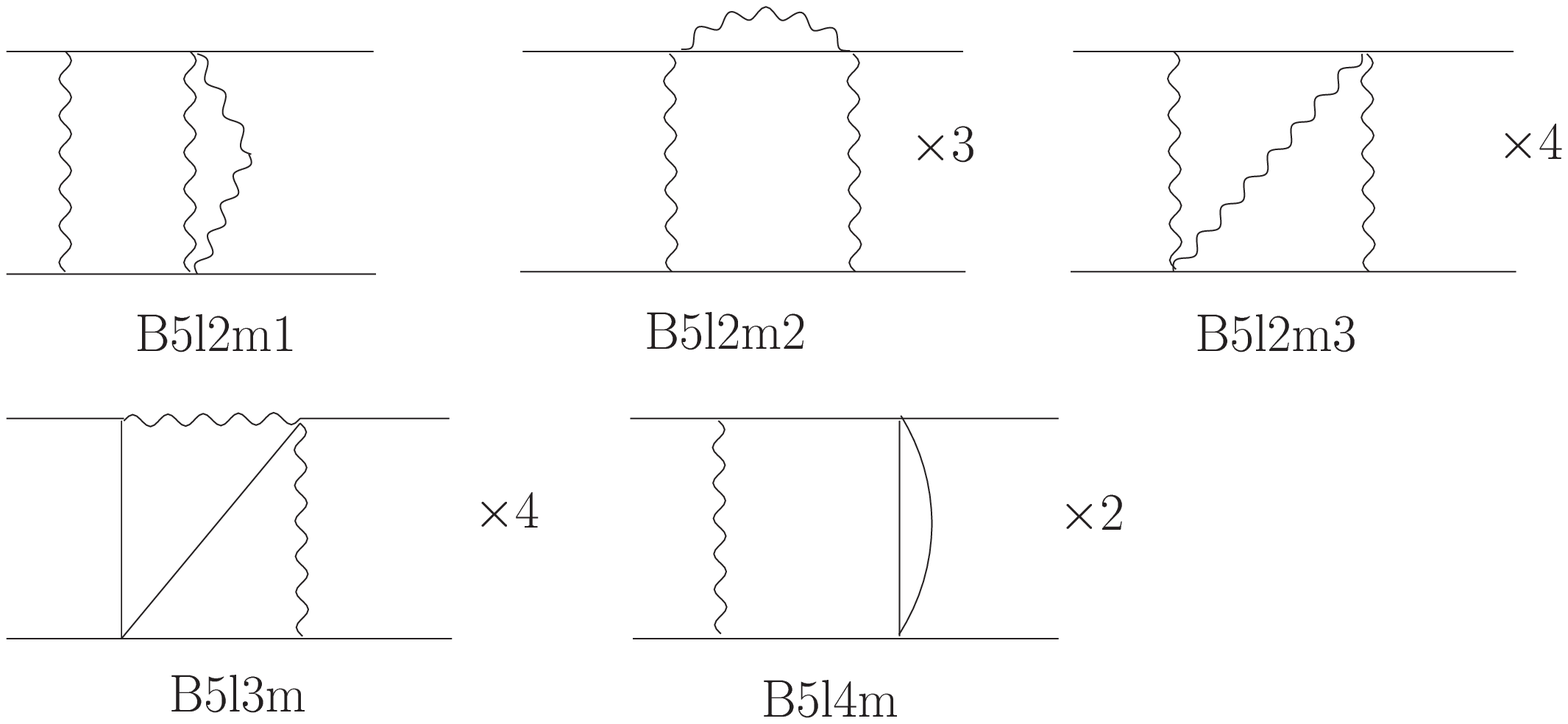, width=7.5cm}
\vspace{-1cm}
\caption{The thirty three 2-loop box MIs (the multiplication factors as in 
Fig. \ref{LL7}.)}
\label{LL3}
\end{figure}

Following the Laporta-Remiddi approach \cite{Laporta:1996mq,Laporta:2001dd}, 
we have determined a set of scalar MIs for the virtual
2-loop corrections to Bhabha scattering.
We use the C++ library
{\tt DiaGen/IdSolver}\footnote{%
{\tt DiaGen}  has already
been used in several other projects 
\cite{Czakon:2002wm,Awramik:2002wn}%
, {\tt IdSolver} is a new software package by
  M.C.
We are also using Fermat \cite{Lewis:2004nn}, FORM
\cite{Vermaseren:2000nd}, Maple, Mathematica.
}%
, which 
determines and solves an appropriate set of algebraic equations,
derived with integration by parts (IBP) \cite{Chetyrkin:1981qh}
and Lorentz invariance (LI) identities \cite{Gehrmann:1999as}.
The latter have been useful for algorithmic optimization 
but did not reduce the number of MIs.  
In Table \ref{table} we give a list of the net numbers of
2-loop master integrals
needed for the evaluation of the various 2-loop vertex and box diagrams.
Obviously, the 1-loop masters add up.
A complete list of 2-loop MIs  is given in Figs. \ref{LL7} to  \ref{LL3}.

%%%%%%%%%%%%%%%%%%%%%%%%%%%%%%%%%%%%%%%%%%%%%%%%%%%%%%%%%%%%%%%%%%%%%%%
\begin{table*}
\setlength{\tabcolsep}{0.8pc}
%%%\newlength{\digitwidth} \settowidth{\digitwidth}{\rm 0}
%%%\catcode`?=\active \def?{\kern\digitwidth}
% -----------------------------------------------------
\caption{The number of 2-loop master integrals needed to calculate the
  2-loop vertex diagrams (Fig. \ref{beta}) and box diagrams (Fig. \ref{LL2})}
\label{table}
%\begin{tabular*}{\textwidth}{@{}l@{\extracolsep{\fill}}rrrr}
%-----------
\begin{tabular}{lrrrrrrrrrrr}
\hline
Diagram           &V1&V2&V3&V4&V5 &B1 &B2 &B3 &B4 &B5 &B6  \\   
\hline
%Master integrals    &&&&&&23 &42 &52 &35 &14 &15  \\
2-point MIs  & 3 & 4  & 4 & 1 & 3 & 4  &5  &5  &4  &3  &3  \\
3-point MIs  & 4 & 10 & 5 & 2 & 1 & 7  &11 &13 &10 &4  &4  \\
Box type MIs    &&&&&&              9  &15 &22 &11 &2  &3  \\
\hline
\end{tabular}
%\end{tabular*}
\end{table*}
%%%%%%%%%%%%%%%%%%%%%%%%%%%%%%%%%%%%%%%%%%%%%%%%%%%%%%%%%%%%%%%%%%%%%%%
%from kinem.tex

\vspace{-.7cm}

\vspace*{-0.7cm}

Strictly speaking, we show prototypes, i.e. 
topologies with masses assigned to the lines.
%diagrams representing
%master integrals with a definite set of denominators. 
Some of the prototype diagrams are multiplied by an integer. 
This indicates that masters appear with dotted lines 
(higher powers of propagators)\footnote{One might 
also use irreducible numerators. However, this introduces an
explicit
dependence of the master on the specific momentum distribution chosen.}.
As an example, we give the master list for the prototype {\tt V5l2m}
(i.e. {\it V}-ertex with {\it 5l}-ines, among them {\it 2m}-assive). 
All the other masters may be found in \cite{web-masters:2004nn}. 
There, we also give 
tables indicating which MIs contribute to SE1-SE4, V1-V5, B1-B6.


We have started to solve  the MIs in a systematic way
 with the differential equation approach proposed in 
\cite{Kotikov:1991hm}
%,Kotikov:1991kg,Kotikov:1991pm,Remiddi:1997ny}
 and developed in \cite{Remiddi:1997ny}\footnote{%
We put the results obtained so far in  
the {\tt Mathematica} file {\tt MastersBhabha.m} in
\cite{web-masters:2004nn}. 
Recalculating 
some of the known results, we found some
misprints in   
Eqs. (80), (85), (121), (128), (131), (144) in 
\cite{Bonciani:2003te} 
and  in Eqs. (36), (38), (39),(B.9) in \cite{Bonciani:2003cj} for
the corresponding masters. 
We have been informed by the authors that an Erratum  is in preparation. 
}.
With $(p_1+p_2)=(p_3+p_4)$ and $p_i^2=m_e^2$, we use
%\begin{eqnarray}
$s= (p_1+p_2)^2,\;t=(p_1-p_3)^2$.
%\end{eqnarray}
The analytical results depend on the variables $x$ and $y$,
\begin{eqnarray}
x = \frac{\sqrt{-s+4}-\sqrt{-s}}{\sqrt{-s+4}+\sqrt{-s}},
%\\
%y = \frac{\sqrt{-t+4}-\sqrt{-t}}{\sqrt{-t+4}+\sqrt{-t}}
\end{eqnarray}

and $y$ is obtained by replacing $s$ by $t$.   
%In the Euclidean region, both $s$ and $t$ are negative, and 
We set everywhere the electron mass to unity, $m_e=1$.

Our normalization of the momentum integrals is chosen such that the
one-loop tadpole is ($d=4-2 \epsilon$):
\begin{eqnarray}
{\tt T1l1m} 
&=& \frac{(\pi e^{\gamma_E})^{\epsilon}}{i\pi^2} 
\int \frac{d^{d}q}{q^2-1} 
\nonumber 
\\
&=& \frac{1}{\epsilon} +1
+\left(1+ \frac{\zeta_2}{2}\right)  \epsilon 
+\ldots
\end{eqnarray}

%-----------------------------------------------------------------------
\section{\label{sec-ver}NEW 2-LOOP 3-POINT MASTERS}
%-----------------------------------------------------------------------
Most of the nineteen 2-loop 3-point MIs needed for the massive Bhabha
process have been calculated already:
{\tt V4l2m} (2 masters),  {\tt V4l3m} (2 masters), {\tt V4l4m} (2 masters), 
{\tt V5l3m} (2 masters), {\tt V6l4m1} (2 masters)  in \cite{Bonciani:2003te}; 
{\tt V4l1m} (4 masters) and {\tt V6l4m2}  in 
 \cite{Bonciani:2003hc}. 
The last four masters completing the list, needed for box diagrams
B1 and B3, are of the type {\tt V5l2m} (see Fig. \ref{LL4}).
We obtain the following analytical expressions: 

\vspace{-0.5cm}

{\allowdisplaybreaks

%\begin{eqnarray}
%{\tt V5l2m1}
% = 
%\frac{x}{(-1 + x^2)}
%  \biggl[ -\zeta_2^2 + \zeta_2H(0, 0, x) \hspace{.5cm} \nonumber \\
% - 2\zeta_2H(1, 0, x) + 
%  2H(0, 0, 0, 1, x) \nonumber \\
% + H(0, 0, 1, 0, x) + 2H(0, 0, 1, 1, x) \nonumber \\
% + 
%  2H(0, 1, 0, 0, x) 
% - 3H(1, 0, 0, 0, x) \nonumber \\
% + 2H(1, 0, 0, 1, x) + 
%  2H(1, 0, 1, 0, x) \nonumber \\
% + 4H(1, 0, 1, 1, x) 
%%\nonumber \\
% - 4 \zeta_3 H(0, x) - 
%      2 \zeta_3 H(1, x) \biggr]
% \nonumber \\
%+ \cal{O}(\epsilon) ,
%\end{eqnarray}
\begin{eqnarray}
{\tt V5l2m1}
 = 
\frac{x}{(-1 + x^2)}
  \biggl[ -\frac{5}{2}\zeta_4 + \zeta_2H(0, 0, x) \hspace{.5cm} 
\nonumber \\
 - 2\zeta_2H(1, 0, x) + 
  2H(0, 0, 0, 1, x) \nonumber \\
 + H(0, 0, 1, 0, x) + 2H(0, 0, 1, 1, x) \nonumber \\
 + 
  2H(0, 1, 0, 0, x) 
 - 3H(1, 0, 0, 0, x) \nonumber \\
 + 2H(1, 0, 0, 1, x) + 
  2H(1, 0, 1, 0, x) \nonumber \\
 + 4H(1, 0, 1, 1, x) 
%\nonumber \\
 - 4 \zeta_3 H(0, x) - 
      2 \zeta_3 H(1, x) \biggr]
 \nonumber \\
+ \cal{O}(\epsilon) ,
\end{eqnarray}

\vspace{-.6cm}

\begin{eqnarray}
{\tt V5l2m1d} 
=  
- \frac{2x}{(-1 + x)^2(1 + x) }
% \nonumber \\
 \biggl[ xH(0, 0, x) \nonumber \\ + 
  (-1 + x)\left[2\zeta_2 
% \nonumber \\
+ H(0, 1, x)\right] \biggr]
% \nonumber \\
+ \cal{O}(\epsilon) ,
\end{eqnarray}

\vspace{-.6cm}
%-----------------------------------------------------------------------

%-----------------------------------------------------------------------
\begin{eqnarray}
{\tt V5l2m2} 
=  
\frac{2x}{(-1 + x^2)}
%\nonumber \\
%%%%%%%%% \biggl[ 16/5 \zeta_2^2 
\biggl[ 8 \zeta_4 
+ 2\zeta_2H(0, 0, x) \nonumber \\
- 4\zeta_2H(0, 1, x) + 
  H(0, 0, 0, 0, x)\nonumber \\
 + 2H(0, 0, 0, 1, x) \biggr]
+ \cal{O}(\epsilon) ,
\end{eqnarray}
%-----------------------------------------------------------------------

\vspace{-.6cm}

\begin{eqnarray}
{\tt V5l2m2d} 
=  -\frac{1}{\epsilon^2}~\frac{x}{4(-1 + x)^2} \hspace{2.5cm}
%-----------------------------------------------------------------------
\nonumber \\
  -~\frac{1}{\epsilon}~\frac{x}{2(-1 +x)^2}\biggl[ -1 
%\nonumber \\
+ H(0, x) 
+ 2H(1, x) \biggr]
\nonumber \\
+~~\frac{x}{4(-1 + x)^2(1 + x)}
\biggl[ -4 - 15\zeta_2 - 4x  
 \nonumber \\
+ \zeta_2x %\nonumber \\
+ 4(1 + x)\left[ H(0, x) + 2H(1, x) \right]  
\nonumber \\
%- 6H(0, 0, x) - 2xH(0, 0, x)  \nonumber \\
-2(x+3) \left[ H(0, 0, x) +2 H(0, 1, x)  \right]
 \nonumber \\
- 8(1+x) \left[ H(1, 0, x)  +2H(1, 1, x) \right] \biggr]
+ \cal{O}(\epsilon) .
\end{eqnarray}
}%end-of-{\allowdisplaybreaks

The $H(0,\ldots,x)$ etc. are harmonic polylogarithms as 
introduced in \cite{Remiddi:1999ew}. 
We have performed a variety of cross-checks on the results, e.g. 
numerical evaluations at fixed kinematical points in the Euclidean
region using sector decomposition
%\cite{Czakon:2004vv},
following the approach described in \cite{Binoth:2000ps}.
The numerical integrations are done with {\tt CUBA} \cite{Hahn:2004fe}.
See also \cite{Gluza:2004gg} for further cross checks.
%\vspace{-2cm}

%-----------------------------------------------------------------------
\section{\label{sec-box}A NEW MASSIVE 2--LOOP BOX MI}
%-----------------------------------------------------------------------
As mentioned, there are thirty three 2-loop box MIs needed.
By now, few of them are known analytically.
For diagram B1, the masters {\tt B7l4m1} and {\tt B7l4m1N} are  given
in \cite{Smirnov:2001cm}; 
for diagram B2, the master {\tt B7l4m2} is known as
two-dimensional integral representation \cite{Smirnov:2004nn};
for  diagram B3, the leading
divergent part of  master {\tt B7l4m3} is published \cite{Smirnov:2004nn}.
For diagram B5, the two double
box masters, {\tt B5l4m} and  {\tt B5l4md},  
have been derived recently with the restriction to $N_f=1$
(one flavor) in
\cite{Bonciani:2003cj,web-masters:2004nn}$^5$.

A  further double box master integral, {\tt B5l2m1} (see Fig. \ref{LL3}), 
is given here; it contributes to
diagrams B1 and B3 and is the solution of the following differential
equation:

\begin{figure}[ht]
\epsfig{file=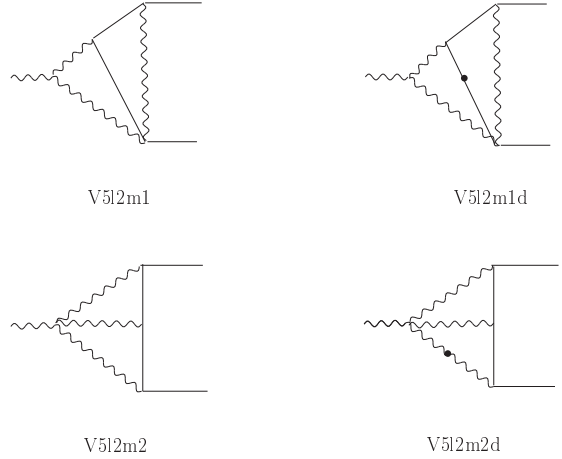, width=7.5cm}
\vspace{-.75cm}
\caption{Four MIs of type V5l2m.}
\label{LL4}
\end{figure}
%\vspace*{2.0cm}

\vspace*{-1.0cm}

\begin{eqnarray}
s \frac{\partial}{\partial s}\; {\tt B5l2m1(s,t)}
= \hspace{3.5cm} \nonumber \\
\biggl[ \frac{8+s^2-2t+s(-6+t+\epsilon t)}{(-4+s)(-4+s+t)} \biggr]  
% \nonumber \\ \times 
\;{\tt B5l2m1(s,t)}
 \nonumber \\
+
\biggl[ \frac{2-9\epsilon+9\epsilon^2}{\epsilon(-4+s)t} \biggr] \;
  {\tt SE3l0m(t)}
 \nonumber \\
+
\biggl[ \frac{(-1+3\epsilon)(-4+t)}{(-4+s)(-4+s+t)} \biggr] \;  {\tt V4l1m2(t)}
 \nonumber \\
-
\biggl[ \frac{s(1-3\epsilon)}{(-4+s)(-4+s+t)} \biggr] \; {\tt V4l2m1(s)}.
\end{eqnarray}

To get this equation we use a simple representation for differential operators.
In the s-channel (for the t-channel operator: $p_2 \leftrightarrow -p_3$)

\begin{eqnarray}
s \frac{\partial }{\partial s}  = \frac12 \left\{
\left(p_1^\mu +  p_2^\mu\right) + \frac{s\left( p_2^\mu -p_3^\mu
  \right)}{s+t-4 m^2}  
    \right\} \frac{\partial}{\partial p_2^\mu} .
\end{eqnarray}

The boundary condition\footnote{We have also used 
the point $x = -1$ or on-shell conditions with cross checks with the 
{\tt ONSHELL2} package \cite{Fleischer:1999tu}.} is fixed at $x=1$.

The solution is:
{\allowdisplaybreaks
\begin{eqnarray}
%B5l2m1[-2,x_] = (x*H(0, x))/(-1 + x^2);
{\tt B5l2m1}
= \frac{1}{\epsilon^2}~\frac{x}{(-1 + x^2)} H(0, x)
  \nonumber \\
 -~~\frac{1}{\epsilon}~\frac{x}{(-1 + x^2)}
\biggl[ \zeta_2 
- H(0, x)\left[ 2 + H(0, y) \right.  \nonumber \\  \left.
+ 2H(1, y) \right]  
+   2H(-1, 0, x) - H(0, 0, x) \biggr] 
 \nonumber \\
+~~ 
%\mbox{Const}+ \cal{O}(\epsilon) 
%\end{eqnarray}
%where
%\begin{eqnarray}
%\mbox{Const} =
%%%%%%%% s \frac{x}{(-1 + x^2)}
           \frac{x}{(-1 + x^2)}
\biggl[ -G(-y, x) \left[ 3\zeta_2 \right. 
 \nonumber \\ \left.
 + H(0, 0, y) + 2H(0, 1, y) \right] + 
  G(-1/y, x) \left[ 5\zeta_2 \right.
 \nonumber \\
\left. + H(0, 0, y) + 2H(0, 1, y) \right] 
-2\zeta_2  \nonumber \\
+ 2G(-1/y, -1, 0, x) 
- G(-1/y, 0, 0, x)  
 \nonumber \\
+  2G(-y, -1, 0, x) 
- G(-y, 0, 0, x) \nonumber \\
+ 2\zeta_2H(-1, x) 
+   4H(0, x) 
- 4\zeta_2H(0, y) 
 \nonumber \\
+ G(-1/y, 0, x)H(0, y) + 
  G(-y, 0, x)H(0, y) 
 \nonumber \\
+ 2H(0, x)H(0, y) + 
  2G(-1/y, 0, x)H(1, y) 
 \nonumber \\
+ 2G(-y, 0, x)H(1, y) + 
  4H(0, x)H(1, y) 
 \nonumber \\
- 4H(-1, 0, x) 
- 2H(0, y)H(-1, 0, x) 
 \nonumber \\
-   4H(1, y)H(-1, 0, x) 
+ 2H(0, 0, x) 
 \nonumber \\
+ 2H(0, x)H(0, 0, y) + 
  4H(0, x)H(0, 1, y) 
 \nonumber \\
+ 4H(0, x)H(1, 0, y) + 
  8H(0, x)H(1, 1, y) 
 \nonumber \\
%%%%%%% + 2H(-1, -1, 0, x) 
        + 4H(-1, -1, 0, x) 
- 2H(-1, 0, 0, x) 
 \nonumber \\
-   4H(0, -1, 0, x) 
+ 2H(0, 0, 0, x)  \nonumber \\
- H(0, 0, 0, y) 
-   2H(0, 0, 1, y) 
- 2\zeta_3 \biggr] + \cal{O}(\epsilon).
%-----------------------------------------------------------------------
\end{eqnarray}
}%end-of-\allowdisplaybreaks

The 2-dimensional harmonic polylogarithms $G$ are introduced in
Appendix A of \cite{Gehrmann:2000zt}.
Here, we use the notations of \cite{Bonciani:2003cj};
the  $G(-y, 0, x)$ and $G(-y, -1, 0, x)$ may be directly taken
from (C.27) and (C.31).
Further it is:
\begin{eqnarray}
G(-y,0,0,x) 
\equiv \frac{1}{2} \int_0^x \frac{dv\ln^2v}{v+y} 
\nonumber \\
= 
\frac{\ln^2x}{2}\ln(1+\frac{x}{y})+\ln{x}
\text{Li}_2(-\frac{x}{y})-\text{Li}_3(-\frac{x}{y}) .
\end{eqnarray}

To summarize, the next, 
certainly ambitious task is the calculation of the remaining
box type  master integrals. 
Finally, for applications at a linear collider,
the complete 2-loop QED amplitudes, the 1-loop Standard Model corrections
\cite{Lorca:2004LL,Fleischer:2002wa}, and cut dependent real photon
radiation effects (determined with Monte Carlo generators) must be evaluated.

%%\input{procbbl_240504}

%\bibliographystyle{/usr1/scratch/riemann/tex/bib/h-elsevier2.bst}
%\bibliography{/afs/ifh.de/group/theorie/tord/bhabha/riemann/tex/bib/2loops}

%%
%%\clearpage
%%
%%\onecolumn
%%
%%The next pages are not part of the proceedings contribution. 
%%
%%\clearpage
%%
%%\input{../../Bhabha/webpages/For_export.tex}

\end{document}